# Robust Key Agreement Schemes

Terence Chan*, Ning Cai† and Alex Grant*
*Institute for Telecommunications Research, University of South Australia
†Xidian University, China

*Abstract*—This paper considers a key agreement problem in which two parties aim to agree on a key by exchanging messages in the presence of adversarial tampering. The aim of the adversary is to disrupt the key agreement process, but there are no secrecy constraints (i.e. we do not insist that the key is kept secret from the adversary). The main results of the paper are coding schemes and bounds on maximum key generation rates for this problem.

## I. INTRODUCTION

In many distributed collaborative algorithms or applications, it is required that each involved party shares a common random key or seed. For instance, in authentication [1] or secret communications [2], the client and the server may need to share a common private key. In another scenario, a common random seed may need to be shared by a group of cooperative users to run a distributed probabilistic algorithm. In such cases, key secrecy may not be important. In all of these examples however, it is important that each party has the *same* key.

It is important to investigate methods for generation and distribution of random keys. For example, in one scenario, it may be required to "divide" a secret key into smaller pieces, for distribution to a group of users. The goal is to ensure that only legitimate groups of users, each of which holds one small piece of the secret, can reconstruct the secret key. This is the secret sharing problem [3]. In another scenario, two legitimate parties (and possibly an adversary) may observe correlated randomness. The objective is for the two parties to extract a common random key from their observations, by exchanging messages over a public channel. The goal is to ensure that an adversary who observes all the messages exchanged over the public channel has no knowledge about the agreed key [4].

The focus of this paper is on robust key agreement in the presence of adversarial tampering (i.e. the adversary can alter some of the messages exchanged between the legitimate parties during the key agreement process). We are interested in coding methods and key generation rates, where the only requirement is that the parties obtain the same key. We do not require that the key be kept secret from the adversary.

One approach to this problem is for one party to simply generate a key and then send it to the other party. Using this simple approach, the key agreement problem reduces to the standard problem of reliable communication. To ensure the other party can reliably reconstruct the key in the presence of noise or tampering, the sender adds redundancy in the form of an error correction code [5]. Recently, the error correction problem was studied in the context of network coding [6]. Although this direct communications approach is very simple, we shall see that its application to key agreement can be suboptimal.

The organization of this paper is as follows. Section II provides the problem formulation. Section III focuses on zero-error key agreement, which is the worst case scenario assuming adversaries have unbounded computational abilities. Section IV considers a weaker adversarial model in which adversaries can only make certain kinds of simple attacks.

*Notations:* Vectors will be denoted by bold-faced lowercase letters whose entries are denoted by superscripts. For example $\mathbf{x}$ is the vector, $x^1$ is its first entry and $x^{[i,j]}$ is the $i^{th}$ to $j^{th}$ entries of $\mathbf{x}$. In addition, define $A_m(n,d)$ as a maximum rate $2^m$-ary code of length $n$ and minimum Hamming distance $d$.

## II. PROBLEM FORMULATION

Consider a simple two-way network as depicted in Figure 1. Alice and Bob aim to agree on a common random key by exchanging messages through the network. Eve is the adversary in the network, whose only objective is to prevent Alice and Bob from agreeing on a key. She attacks by replacing some of the exchanged messages. There is no requirement to keep the key secret from Eve.

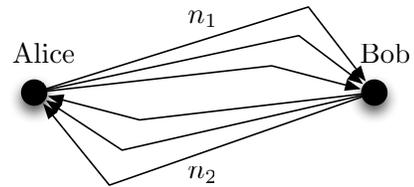

Fig. 1. A two-way network

We will mainly consider a two-round key agreement scenario. In the first round, Alice generates $n_1$ message vectors, $\mathbf{x}_1, \ldots, \mathbf{x}_{n_1}$. Assume without loss of generality that all messages are binary vectors of length $m$. These are sent to Bob using the $n_1$ forward links (one for each message). Eve observes the messages and can replace some of them with messages of her own choosing. Let the $n_1$ messages received by Bob be denoted $\hat{\mathbf{x}}_1, \ldots \hat{\mathbf{x}}_{n_1}$.

In the second round, after receiving $\hat{\mathbf{x}}_1, \ldots \hat{\mathbf{x}}_{n_1}$, Bob generates $n_2$ message vectors, $\mathbf{y}_1, \ldots, \mathbf{y}_{n_2}$. These are sent to Alice using the $n_2$ backward links. Again, Eve may observe and replace some of the messages. Let the $n_2$ messages received by Alice be denoted $\hat{\mathbf{y}}_1, \ldots \hat{\mathbf{y}}_{n_2}$.

If Eve can attack every link, it is impossible for Alice and Bob to agree on a key. However there are many scenarios of interest where it may be reasonable to assume that this is not possible (e.g. due to limited network access, or the use of special hardened links). Henceforth, we assume that Eve can attack at most $t$ links in total. In other words,

$$\sum_{i=1}^{n_1} d_H(\mathbf{x}_i, \hat{\mathbf{x}}_i) + \sum_{i=1}^{n_2} d_H(\mathbf{y}_i, \hat{\mathbf{y}}_i) \leq t \quad (1)$$

where $d_H(\cdot)$ is a Hamming distortion function with

$$d_H(\mathbf{x}, \mathbf{y}) = \begin{cases} 0 & \mathbf{x} = \mathbf{y} \\ 1 & \mathbf{x} \neq \mathbf{y} \end{cases}$$

i.e. two distinct vectors are at distance 1, regardless of how many element disagree.

After these two rounds of message exchange (forward and backward), Alice and Bob make independent decisions on their random key. Let $g_a$ and $g_b$ be their (key-)decoding functions respectively. A key agreement error occurs if

$$g_a(\mathbf{x}_1, \ldots, \mathbf{x}_{n_1}, \hat{\mathbf{y}}_1, \ldots \hat{\mathbf{y}}_{n_2}) \neq g_b(\mathbf{y}_1, \ldots, \mathbf{y}_{n_2}, \hat{\mathbf{x}}_1, \ldots \hat{\mathbf{x}}_{n_1}).$$

A key agreement scheme is specified by the encoders and decoders used by Alice and Bob. We shall use a probabilistic setting. Alice's encoder $E_a$ is specified by a probability distribution $\Pr(\mathbf{x}_1, \ldots, \mathbf{x}_{n_1})$ which governs how Alice generates the first round of messages. Bob's encoder $E_b$ however is specified by a conditional distribution $\Pr(\mathbf{y}_1, \ldots, \mathbf{y}_{n_2} \mid \hat{\mathbf{x}}_1, \ldots \hat{\mathbf{x}}_{n_1})$ which determines how the second round of messages should be generated after receiving the possibly corrupted messages from Alice. A key agreement scheme will be denoted by the tuple $(m, n_1, n_2, E_a, E_b, g_a, g_b)$ or simply $(E_a, E_b, g_a, g_b)$ if $m, n_1, n_2$ are understood.

Eve's attack is specified by a pair of conditional probability distributions

$$\Pr(\hat{\mathbf{x}}_1, \ldots \hat{\mathbf{x}}_{n_1} | \mathbf{x}_1, \ldots, \mathbf{x}_{n_1}) \quad (2)$$

and

$$\Pr(\hat{\mathbf{y}}_1, \ldots \hat{\mathbf{y}}_{n_2} | \mathbf{y}_1, \ldots, \mathbf{y}_{n_2}, \mathbf{x}_1, \ldots, \mathbf{x}_{n_1}, \hat{\mathbf{x}}_1, \ldots, \hat{\mathbf{x}}_{n_1}), \quad (3)$$

These distributions must satisfy the constraint (1). Let

$$K_1 = g_a(\mathbf{x}_1, \ldots, \mathbf{x}_{n_1}, \hat{\mathbf{y}}_1, \ldots \hat{\mathbf{y}}_{n_2})$$
$$K_2 = g_b(\mathbf{y}_1, \ldots, \mathbf{y}_{n_2}, \hat{\mathbf{x}}_1, \ldots \hat{\mathbf{x}}_{n_1}).$$

The probability distribution of $K_1$ and $K_2$ depends on Eve's attacking strategy. For a given attacking strategy $E$, let

$$H_E(K_1 | K_1 = K_2)$$
$$\triangleq -\sum_{k_1} \Pr(K_1 = k_1 | K_1 = K_2) \log \Pr(K_1 = k_1 | K_1 = K_2)$$

where $\Pr(K_1 = k_1 | K_1 = K_2)$ is the conditional probability that $K_1 = k_1$ given the event that $K_1 = K_2$.

Let $\mathcal{A}_E$ be the set of attacking strategies that Eve can choose (i.e. the set of pairs of conditional distributions (2) and (3) satisfying (1)). We define the key agreement rate (for a given key agreement scheme) as

$$\min_{E \in \mathcal{A}_E} H_E(K_1 | K_1 = K_2).$$

### III. ZERO-ERROR KEY AGREEMENT

The objective of zero-error key agreement is for Alice and Bob to generate identical keys with probability one at some positive rate.

*Definition 1:* For given positive integers $n_1, n_2, m$, the key rate $R$ is called zero-error admissible if there exists a key agreement scheme $(E_a, E_b, g_a, g_b)$ such that (1) the probability of key agreement error is zero for all attacking strategies that Eve can choose and (2) $R \leq \min_{E \in \mathcal{A}_E} H_E(K_1)$. The zero-error key agreement capacity is the supremum of all zero-error admissible rates.

The natural fundamental question is: What is the zero-error key agreement capacity? In this paper, we will give lower bounds for the zero-error key agreement capacity and simple schemes that achieve the lower bounds.

*Theorem 1:* If $t \geq \max(n_1, n_2)$, then the zero-error key agreement capacity is 0.

*Proof sketch:* Since $t \geq n_1, n_2$, no matter which messages Alice and Bob send, Eve can replace them with any other messages. If the probability of key agreement error is zero, then the key that Alice and Bob agree on must be independent of $\hat{\mathbf{x}}_1, \ldots \hat{\mathbf{x}}_{n_1}$ and $\hat{\mathbf{y}}_1, \ldots \hat{\mathbf{y}}_{n_2}$. As such, the agreed key can only be a constant. ∎

#### A. Examples

We will now develop some small examples that provide motivation for a general coding scheme.

If messages can be sent only in one direction (i.e., either $n_1$ or $n_2$ is zero), then key agreement is equivalent to transmission of a random key from one party to another. When messages can be sent in both directions, we can naively decouple the two rounds of message transmissions into two rounds of random key transmissions as follows.

*Example 1 (Direct key transmission):* Suppose $n_1 = n_2 = 3$ and $t = 1$. Let $\mathcal{C}_a = \{(\mathbf{x}_1, \mathbf{x}_2, \mathbf{x}_3) : \mathbf{x}_1 = \mathbf{x}_2 = \mathbf{x}_3\}$ and let

$$\Pr(\mathbf{x}_1, \mathbf{x}_2, \mathbf{x}_3) \triangleq \begin{cases} 1/2^m & \text{if } (\mathbf{x}_1, \mathbf{x}_2, \mathbf{x}_3) \in \mathcal{C}_a \\ 0 & \text{otherwise.} \end{cases}$$

Since $\mathcal{C}_a$ has minimum distance 3, no matter how Eve attacks, Bob can reconstruct $\mathbf{x}_1$ without error. Note that, if the minimum distance of $\mathcal{C}_a$ is less than 3, then Bob may fail to correctly reconstruct $\mathbf{x}_1$.

Similarly, let $\mathcal{C}_b = \{(\mathbf{y}_1, \mathbf{y}_2, \mathbf{y}_3) : \mathbf{y}_1 = \mathbf{y}_2 = \mathbf{y}_3\}$ and

$$\Pr(\mathbf{y}_1, \mathbf{y}_2, \mathbf{y}_3 | \hat{\mathbf{x}}_1, \hat{\mathbf{x}}_2, \hat{\mathbf{x}}_3) \triangleq \begin{cases} 1/2^m & \text{if } (\mathbf{y}_1, \mathbf{y}_2, \mathbf{y}_3) \in \mathcal{C}_b \\ 0 & \text{otherwise} \end{cases}$$

for all $(\hat{\mathbf{x}}_1, \hat{\mathbf{x}}_2, \hat{\mathbf{x}}_3)$. Again, Alice can reconstruct $\mathbf{y}_1$ without error, no matter how Eve attacks. Finally, Alice and Bob can use $(\mathbf{x}_1, \mathbf{y}_1)$ as the common random key whose entropy is $2m$.

The above scheme essentially consists of two one-round key transmission schemes. The resulting key consists of two random parts, one generated by Alice (and sent to Bob) and one generated by Bob (and sent to to Alice). Despite its simplicity, this scheme is not optimal as shown by the following example.

*Example 2:* Suppose $n_1 = n_2 = 3$ and $t = 1$. Let $\mathcal{C}_a$ be an $A_m(3, 2)$ code and

$$\Pr(\mathbf{x}_1, \mathbf{x}_2, \mathbf{x}_3) \triangleq \begin{cases} 1/|\mathcal{C}_a| & \text{if } (\mathbf{x}_1, \mathbf{x}_2, \mathbf{x}_3) \in \mathcal{C}_a \\ 0 & \text{otherwise.} \end{cases}$$

Note that $\mathcal{C}_a$ has minimum distance 2. Therefore, if Eve attacks one of the forward links, Bob can always detect it but not necessarily correct it.

Consider the following codebooks $\mathcal{C}_{b,0}^* = A_{m-1}(3, 3)$ and $\mathcal{C}_{b,1}^* = A_{m-1}(3, 1)$. Let

$$\mathcal{C}_{b,0} = \left\{ \begin{array}{c} (\mathbf{y}_1, \mathbf{y}_2, \mathbf{y}_3) : y_1^1 = y_2^1 = y_3^1 = 0 \\ \text{and } (y_1^{[2,m]}, y_2^{[2,m]}, y_3^{[2,m]}) \in \mathcal{C}_{b,0}^* \end{array} \right\},$$

$$\mathcal{C}_{b,1} = \left\{ \begin{array}{c} (\mathbf{y}_1, \mathbf{y}_2, \mathbf{y}_3) : y_1^1 = y_2^1 = y_3^1 = 1 \\ \text{and } (y_1^{[2,m]}, y_2^{[2,m]}, y_3^{[2,m]}) \in \mathcal{C}_{b,1}^* \end{array} \right\}$$

If Bob does not detect any errors (i.e., $(\hat{\mathbf{x}}_1, \hat{\mathbf{x}}_2, \hat{\mathbf{x}}_3) \in \mathcal{C}_a$), then

$$\Pr(\mathbf{y}_1, \mathbf{y}_2, \mathbf{y}_3 | \hat{\mathbf{x}}_1, \hat{\mathbf{x}}_2, \hat{\mathbf{x}}_3) \begin{cases} 1/|\mathcal{C}_{b,0}| & \text{if } (\mathbf{y}_1, \mathbf{y}_2, \mathbf{y}_3) \in \mathcal{C}_{b,0} \\ 0 & \text{otherwise.} \end{cases}$$

Otherwise, if an error is detected,

$$\Pr(\mathbf{y}_1, \mathbf{y}_2, \mathbf{y}_3 | \hat{\mathbf{x}}_1, \hat{\mathbf{x}}_2, \hat{\mathbf{x}}_3) \triangleq \begin{cases} 1/|\mathcal{C}_{b,1}| & \text{if } (\mathbf{y}_1, \mathbf{y}_2, \mathbf{y}_3) \in \mathcal{C}_{b,1} \\ 0 & \text{otherwise.} \end{cases}$$

After receiving $\hat{y}_1^1, \hat{y}_2^1, \hat{y}_3^1$, Alice can reconstruct $y_1^1$. Therefore, Alice will know which codebook Bob used. It is easy to see that $(\mathbf{y}_1, \mathbf{y}_2, \mathbf{y}_3)$ can also be reconstructed perfectly.

Finally, Alice and Bob agree on $K = (k_o, k_a, k_b)$ such that
- $k_o = y_1^1$, which indicates whether errors were detected in the forward links;
- $k_a = 0$ if $k_o = 1$. Otherwise, $k_a = (\mathbf{x}_1, \mathbf{x}_2, \mathbf{x}_3)$;
- $k_b = (\mathbf{y}_1, \mathbf{y}_2, \mathbf{y}_3)$.

It is straightforward to prove that the probability of key agreement error is zero and that the entropy of the key $K$ is at least

$$\min(\log |A_m(3,2)| + \log |A_{m-1}(3,3)|, \log |A_{m-1}(3,1)|).$$

When $m$ is sufficiently large, the Singleton bound is tight, and hence the entropy of $K$ is at least $3m - 1$.

Compared with the key agreement rate in Example 1, a 50% gain is achieved.

From the above example, it is easy to see that the direct key transmission scheme in Example 1 is suboptimal because Bob did not use his received messages to estimate how many forward links were attacked by Eve. As a result, Bob has to pessimistically protect his messages, assuming that Eve can attack $t$ backward links.

Although the key agreement scheme in Example 2 may appear to be a modified direct key transmission scheme, there are some subtle differences. Using direct key transmission (multiple one-round key distribution sessions), the agreed key consists of two random parts, one from Alice and one from Bob. The entropy of the agreed key will be the same no matter how Eve attacks. On the other hand, in the scheme detailed in Example 2, the size of the key depends on how Eve attacks. For instance, if Eve attacks the forward link, then the entropy of the resulting key is the largest. Furthermore, in this case, the key is essentially solely generated by Bob.

In this paper, we are not concerned with the source of randomness. However, in some other scenarios, it may be of a practical concern. For example, suppose that there is another adversary who can "observe" how Bob can generate the random messages $(\mathbf{y}_1, \mathbf{y}_2, \mathbf{y}_3)$. Then, it may cause a problem if that adversary will know the key completely.

The following is another interesting example in which direct key transmission fails altogether, but the key agreement capacity is nonzero.

*Example 3* ($n_1 = n_2 = 2$ and $t = 1$): Let $\mathcal{C}_a = A_m(2, 2)$ and

$$\Pr(\mathbf{x}_1, \mathbf{x}_2) \triangleq \begin{cases} 1/|\mathcal{C}_a| & \text{if } (\mathbf{x}_1, \mathbf{x}_2) \in \mathcal{C}_a \\ 0 & \text{otherwise.} \end{cases}$$

Again, if Eve attacks the forward links, Bob can detect it but not correct it.

Consider codebooks $\mathcal{C}_{b,0}^* = A_{m-1}(2, 2)$ and $\mathcal{C}_{b,1}^* = A_{m-1}(2, 1)$. Let

$$\mathcal{C}_{b,0} = \left\{ \begin{array}{c} (\mathbf{y}_1, \mathbf{y}_2) : y_1^1 = y_2^1 = 0 \\ \text{and } (y_1^{[2,m]}, y_2^{[2,m]}) \in \mathcal{C}_{b,0}^* \end{array} \right\},$$

$$\mathcal{C}_{b,1} = \left\{ \begin{array}{c} (\mathbf{y}_1, \mathbf{y}_2) : y_1^1 = y_2^1 = 1 \\ \text{and } (y_1^{[2,m]}, y_2^{[2,m]}) \in \mathcal{C}_{b,1}^* \end{array} \right\}.$$

If Bob does not detect any errors (i.e., $(\hat{\mathbf{x}}_1, \hat{\mathbf{x}}_2) \in \mathcal{C}_a$), then

$$\Pr(\mathbf{y}_1, \mathbf{y}_2 | \hat{\mathbf{x}}_1, \hat{\mathbf{x}}_2) \begin{cases} 1/|\mathcal{C}_{b,0}| & \text{if } (\mathbf{y}_1, \mathbf{y}_2) \in \mathcal{C}_{b,0} \\ 0 & \text{otherwise.} \end{cases}$$

Otherwise,

$$\Pr(\mathbf{y}_1, \mathbf{y}_2 | \hat{\mathbf{x}}_1, \hat{\mathbf{x}}_2) \begin{cases} 1/|\mathcal{C}_{b,1}| & \text{if } (\mathbf{y}_1, \mathbf{y}_2) \in \mathcal{C}_{b,1} \\ 0 & \text{otherwise.} \end{cases}$$

As before, we can easily show that the resulting key agreement capacity is at least $\log |\mathcal{C}_a| = m$.

### B. Generalization

We will now generalize Examples 2 and 3 to arbitrary $n_1, n_2$ and $t$. Let $\ell = \lceil \log(t+1) \rceil$ and $\Omega_m(d, t_1)^1$ be defined as follows:

$$\Omega_m(d, t_1) = \begin{cases} \log |A_m(n_1, d)| |A_{m-\ell}(n_2, 2t+1)| & d > t + t_1 \\ \log |A_{m-\ell}(n_2, 2(t-t_1)+1)| & \text{otherwise.} \end{cases}$$

*Theorem 2 (Inner bound):* Suppose that $n_2 > 2t$. Then the zero-error key agreement capacity is at least

$$\max_{n_1 \geq d > t} \min(\Omega_m(d, 0), \Omega_m(d, d - t)). \tag{4}$$

---
[1]We do not explicitly indicate the dependency of $\Omega_m(d, t_1)$ on $n_1, n_2, t$ to simplify notations.

*Proof:* Let $\mathcal{C}_a$ be an $A_m(n_1, d)$ code and

$$\Pr(\mathbf{x}_1, \ldots, \mathbf{x}_{n_1}) \triangleq \begin{cases} 1/|\mathcal{C}_a| & \text{if } (\mathbf{x}_1, \ldots, \mathbf{x}_{n_1}) \in \mathcal{C}_a \\ 0 & \text{otherwise.} \end{cases}$$

Let $t_1$ be the number of forward links that Eve attacks. If $t_1 < d - t$, then Bob can reconstruct $(\mathbf{x}_1, \ldots, \mathbf{x}_{n_1})$ perfectly. Otherwise, Bob can deduce that at least $d - t$ forward links have been attacked by Eve.

Any integer $i$ between $0$ and $t$, can be easily represented using $\ell$ bits. For each $i$, let $\mathcal{C}_{b,i}^* = A_{m-\ell}(n_2, 2(t-i)+1)$ and

$$\mathcal{C}_{b,i} \triangleq \left\{ \begin{array}{c} (\mathbf{y}_1, \ldots, \mathbf{y}_{n_2}) : y_1^{[1,\ell]} = \cdots = y_{n_2}^{[1,\ell]} = i \\ \text{and } (y_1^{[\ell+1,m]}, \ldots, y_{n_2}^{[\ell+1,m]}) \in \mathcal{C}_{b,i}^* \end{array} \right\}.$$

Suppose that Bob can detect and correct errors (i.e., $(\hat{\mathbf{x}}_1, \ldots, \hat{\mathbf{x}}_{n_1})$ is within a distance of $d-t-1$ from a codeword in $\mathcal{C}_a$), then he can also determine the number of forward links $i$ that were attacked by Eve. Then let

$$\Pr(\mathbf{y}_1, \ldots, \mathbf{y}_{n_2} | \hat{\mathbf{x}}_1, \ldots, \hat{\mathbf{x}}_{n_1})$$
$$= \begin{cases} 1/|\mathcal{C}_{b,i}| & \text{if } (\mathbf{y}_1, \ldots, \mathbf{y}_{n_2}) \in \mathcal{C}_{b,i} \\ 0 & \text{otherwise.} \end{cases}$$

Similarly, if Bob determines that at least $d-t$ errors occur in the forward links, then

$$\Pr(\mathbf{y}_1, \ldots, \mathbf{y}_{n_2} | \hat{\mathbf{x}}_1, \ldots, \hat{\mathbf{x}}_{n_1})$$
$$= \begin{cases} 1/|\mathcal{C}_{b,d-t}| & \text{if } (\mathbf{y}_1, \ldots, \mathbf{y}_{n_2}) \in \mathcal{C}_{b,d-t} \\ 0 & \text{otherwise.} \end{cases}$$

Let $K = (k_o, k_a, k_b)$ such that
- $k_o = y_1^{[1,\ell]}$ which is the number of errors (or attacks) occurred in the forward links;
- $k_a = 0$ if $k_o = d - t$. Otherwise, $k_a = (\mathbf{x}_1, \ldots, \mathbf{x}_{n_1})$;
- $k_b = (\mathbf{y}_1, \ldots, \mathbf{y}_{n_2})$.

It is straightforward to prove that $K$ is known to both Alice and Bob, and the entropy of the common key $K$ is at least

$$H(K) \geq \min_{0 \leq t_1 \leq t} \Omega_m(d, t_1) \quad (5)$$
$$= \min(\Omega_m(d, 0), \Omega_m(d, d-t)) \quad (6)$$

and the result then follows. ■

In above, we considered only two-round key-agreement schemes and obtained inner bounds on rates of the agreed key. We can easily extend the bounds to multi-round scenarios.

Define $\mathcal{R}_{w,n_1,\ldots,n_w,t,m}$ as the key agreement capacity in a $w$-round key agreement scenario in which (1) the number of messages that can be sent in the $i^{th}$ round is $n_i$, (2) the maximum number of links that can be attacked by Eve is $t$, and (3) each message is a binary vector of length $m$.

*Theorem 3:* Suppose $n_1, \ldots, n_w > 2t+1$. Then for any $d$ such that $2t \geq d > t$, $\mathcal{R}_{w,n_1,\ldots,n_w,t,m}$ is at least

$$\min \left( \begin{array}{c} \log |A_{m-\ell}(n_1, d)| + \mathcal{R}_{w-1,n_2,\ldots,n_w,t,m-\ell}, \\ \mathcal{R}_{w-1,n_2,\ldots,n_w,2t-d,m-\ell} \end{array} \right) \quad (7)$$

where $\ell = \lceil \log(t+1) \rceil$.

**Remark:** By replacing the key agreement capacity terms in (7) with their corresponding inner bounds, we can get inner bounds for the multi-round key agreement capacity from Theorems 2 and 3.

## IV. RANDOM ERRORS

In the previous section, we considered the worst case scenario in which no errors are allowed in key agreement. Even if Eve attacks links randomly, there is still a small but positive probability that she may choose the most damaging attack. In fact, in this worst case scenario, we can even assume that Eve has knowledge the messages sent by Alice prior to attack.

We will now relax our model to allow small errors and assume that it is infeasible for Eve to determine which is the most damaging attack. More specifically, Eve can only decide on the number of links to be attacked in each direction, but not explicitly which links. We will consider the asymptotic case:

$$n_1 = \lambda_1 r, \quad n_2 = \lambda_2 r, \text{ and } t = \tau r$$

where $r$ approaches infinity. Also, each link can transmit either zero or one (i.e., $m = 1$).

*Definition 2:* A normalized key agreement rate $R$ is $\epsilon$-error admissible (with respect to given $\lambda_1, \lambda_2$ and $\tau$) if there exists a sequence of key agreement schemes $(1, n_1, n_2, t, E_a, E_b, g_a, g_b)^2$ such that
1) $\lim_{r \to \infty} n_1/r = \lambda_1$, $\lim_{r \to \infty} n_2/r = \lambda_2$ and $\lim_{r \to \infty} t/r = \tau$,
2) $R \leq \min_{E \in \mathcal{A}_E} H_E(K_1 | K_1 = K_2)/r$,
3) The probability of key agreement failure, denoted as $P_e(1, n_1, n_2, t, E_a, E_b, g_a, g_b)$, goes to zero as $r$ goes to infinity.

The normalized $\epsilon$-error key agreement capacity (for given $\lambda_1, \lambda_2$ and $\tau$) is the supremum $\epsilon$-error admissible $R$. In the following, we will obtain a lower bound for the capacity.

*Definition 3 (Combinatorial Binary Symmetric Channel):* A $CBS(\epsilon)$ channel takes binary inputs and gives binary output. Let $(X_1, \ldots, X_n)$ be the $n$ input symbols to the channel and $(\hat{X}_1, \ldots, \hat{X}_n)$ be the $n$ output symbols. The channel inputs and outputs are related by

$$(\hat{X}_1, \ldots, \hat{X}_n) = (X_1, \ldots, X_n) \oplus (E_1, \ldots, E_n),$$

where $(E_1, \ldots, E_n)$ is a binary error vector, independent of the inputs and uniformly distributed over $\{(e_1, \ldots, e_n) : d_H(e_1, \ldots, e_n) \leq n\epsilon\}$ where $\epsilon < 1/2$ is a channel parameter.

The $CBS(\epsilon)$ channel is not memoryless, but behaves like a memoryless binary symmetric channel with crossover probability $\epsilon$ for sufficiently large $n$.

Let $\mathcal{X} = \{0, 1\}$. Consider a rate $s$ error correcting/detecting code. The encoder is a mapping $f : \{1, \ldots, 2^{ns}\} \mapsto \mathcal{X}^n$ and the decoder is a mapping $g : \mathcal{X}^n \mapsto \{0, 1, \ldots, 2^{ns}\}$ where decoder output zero means that the decoder fails to correct errors. Suppose that the transmitted codeword is $f(i)$.

---
[2]The sequence of schemes are indexed by $r$. For notational simplicity, we do not indicate the dependency explicitly.

A *correction failure* means that decoding output is not $i$ and a *detection failure* means that the output is neither $i$ nor 0.

Clearly, the probabilities of correction and detection failures depend on the channel model. In this paper, we will focus on CBS channels. For a given error correcting/detecting code $\mathcal{C}$, Let $P_e^c(\mathcal{C}, \xi)$ and $P_e^d(\mathcal{C}, \xi)$ be respectively the probabilities of correction and detection failures when the channel is a $CBS(\xi)$ channel.

*Proposition 1 (Achievability):* Fix $\xi < 1/2$. Let $I(\xi) \triangleq 1 + \xi \log \xi + (1-\xi) \log(1-\xi)$. For any $s < I(\xi)$, we can construct a sequence of rate $s_n$ error correcting/detecting codes $\mathcal{C}^n$ such that

$$\liminf_{i \to \infty} s_n \geq s \tag{8}$$

$$\lim_{n \to \infty} P_e^c(\mathcal{C}^n, \epsilon) = 0, \text{for } \epsilon \leq \xi \tag{9}$$

$$\lim_{n \to \infty} P_e^d(\mathcal{C}^n, \epsilon) = 0, \forall \epsilon \tag{10}$$

*Proof:* The sequence of codes $\mathcal{C}^n$ is randomly constructed as follows. The proof of (9) and (10) is straightforward and will be omitted.

The encoder $f$ is a randomly selected mapping

$$f : \{1, \ldots, 2^{ns}\} \mapsto \mathcal{X}^n$$

such that each symbol in the codeword $f(i)$ is independently and uniformly distributed over $\{0, 1\}$.

The decoder $g$ is a "bounded distance decoder"

$$g : \mathcal{X}^n \mapsto \{0, 1, \ldots, 2^{ns}\}.$$

For any sequence $(\hat{X}_1, \ldots, \hat{X}_n)$, if there exists a unique $f(i)$ such that if the Hamming weight of $(\hat{X}_1, \ldots, \hat{X}_n) - f(i)$ is less than $n\xi$, then the decoder output will be $i$. Otherwise, $g(\hat{X}_1, \ldots, \hat{X}_n) = 0$.
∎

Proposition 1 proves the existence of error correcting/detecting codes that can correct $\xi$-fraction of errors. We can use these codes to construct key agreement schemes as before. As a result, we obtain the following bounds on the $\epsilon$-error key agreement capacity.

*Theorem 4 (Inner bound):* Assume $\tau/\lambda_2 < 1/2$. Let $\gamma = (\tau - \lambda_1 \xi)/\lambda_2$. Then the key agreement rate is at least

$$\max_{\xi \leq \tau/\lambda_1} \left( \lambda_2 I(\gamma/\lambda_2), \lambda_1 I(\xi) + \lambda_2 I(\tau/\lambda_2) \right). \tag{11}$$

*Proof:* Let $r\nu$ be the number of links attacked in the forward direction. By Proposition 1, for sufficiently large $r$, there exists a code at rate close to $I(\xi)$ which with high probability can correct any $n_1 \xi = r\lambda_1 \xi$'s errors and detect any number of errors.

That $\tau/\lambda_2 < 1/2$ guarantees that Bob can successfully inform Alice whether he can correctly decode his received message. If Bob's decoder can correct the errors, then he and Alice both know $(X_1, \ldots, X_{n_1})$. Otherwise, Bob can determine that Eve has attacked at least $r\lambda_1 \xi$ links and that she can attack at most $r\gamma \triangleq r\tau - r\lambda_1 \xi$ of backward links. Hence Bob can transmit $r\lambda_2 I(\gamma/\lambda_2)$ bits of random key to Alice.

Depending on the number of forward link attacks made by Eve, the key agreement rate is given by

$$\Omega(\xi, \nu) = \begin{cases} \lambda_1 I(\xi) + \lambda_2 I(\tau/\lambda_2) & \text{if } \nu < \lambda_1 \xi \\ \lambda_2 I(\gamma/\lambda_2) & \text{if } \nu \geq \lambda_1 \xi \end{cases}$$

Alice and Bob can agree on a common random key at rate no less than

$$\max_{\xi \leq \tau/\lambda_1} \min_{0 \leq \nu \leq \tau} \Omega(\xi, \nu) \tag{12}$$

By monotonicity of the function $I(\cdot)$, we can further reduce (12) to (11) and hence the result follows. ∎

*Note:* If both Alice and Bob share a small private key which is unknown by Eve, they can use the private key in a way so that any attacks made by Eve are no better than a random attack.

## V. CONCLUSION

In this paper, we consider a key (or seed) agreement problem in which two parties aim to agree on a key by exchanging messages in the presence of adversarial tampering. We showed that naively decoupling the problem into two key transmission problems is suboptimal. We proposed an improved scheme and obtained lower bounds on the key generation rates. Although the proposed scheme is very simple, it can significantly improve the key agreement rate. Finally, we extended the proposed schemes and bounds to a weaker scenario in which the adversary has a limited computational power and cannot select the most damaging attacks.